# Steam Degradation of Ytterbium Disilicate Environmental Barrier Coatings: Effect of Composition, Microstructure and Temperature

Daniel Tejero-Martin*[1], Mingwen Bai[2], Acacio R. Romero[1], Richard G. Wellman[3], Tanvir Hussain[1*]

[1] *Faculty of Engineering, University of Nottingham, University Park, Nottingham, NG7 2RD*

[2] *Institute for Future Transport and Cities, Coventry University, Priory St, Coventry, CV1 5FB, UK*

[3] *Surface Engineering, Rolls-Royce Plc, Derby, DE24 8BJ*

\* daniel.tejero@nottingham.ac.uk

\* tanvir.hussain@nottingham.ac.uk; +44 115 951 3795

**Abstract**

Recession of environmental barrier coatings (EBC) in environments containing steam is a pressing concern that requires further research before their implementation in gas turbine engines can be realized. In this work, free-standing plasma sprayed $Yb_2Si_2O_7$ coatings were exposed to flowing steam at 1350 °C and 1400 °C for 96 h. Three samples were investigated, one coating with a low porosity level (< 3 %) and 1 wt.% $Al_2O_3$ representing traditional EBCs; and two coatings with higher porosity levels (~20 %) representing abradable EBCs. Phase composition and microstructural evolution were studied in order to reveal the underlying mechanism for the interaction between high temperature steam and ytterbium disilicate. The results show depletion of $Yb_2SiO_5$ near the surface and formation of ytterbium garnet ($Yb_3Al_5O_{12}$) on top of all three coatings due to the reaction with gaseous Al-containing impurities coming from the alumina furnace tubes. The 1 wt.% $Al_2O_3$ added to the EBC sample exacerbated the formation of garnet at 1400 °C compared to the abradable samples, which presented lower quantities of garnet. Additionally, inter-splat boundaries were visible after exposure, indicating preferential ingress of gaseous Al-containing impurities through the splat boundaries.

## 1. Introduction

Nickel-based super-alloys have allowed the current generation of gas turbine engines for aerospace and energy generation to reach extraordinary levels of efficiency. Despite advances in protective coatings and active cooling, the service temperature is ultimately limited to the melting point of the nickel-based super-alloy substrate. SiC/SiC ceramic matrix composites (CMCs) have been identified as a suitable replacement as the material to be used in the hot section for the next generation of gas turbines. Their increased service temperature and superior strength/weight ratio at high temperatures compared to nickel-based super-alloys [1,2] are regarded as the key to improving the performance and weight of future gas turbine engines.

Before nickel-based super-alloys can be effectively replaced with CMCs [3], an effective and reliable protection against corrosion and degradation during service must be developed. CMCs exposed to service conditions face two main degradation mechanisms. Firstly, the presence of calcium magnesium alumina-silicates corrosive species (generally labelled as CMAS for convenience [4–6]) can cause molten deposits that interact with the components, shortening their service life. CMAS can be present by the ingestion of debris during take-off and landing, as well as when flying over arid environments or due to the presence of airborne volcanic ash [7,8]. Secondly, CMCs exposed to high temperatures under clean, dry oxygen form a protective $SiO_2$ scale that provides protection against corrosion and recession [9]. Under the presence of steam, a naturally occurring combustion product [10,11], the CMCs components show increased oxidation [12–14] and accelerated corrosion due to the volatilization of the $SiO_2$ scale to form gaseous Si-O-H species, such as $Si(OH_4)$ [15], as shown below:

$$SiC + 1.5O_2(g) = SiO_2 + CO(g) \tag{1}$$

$$SiO_2 + 2H_2O(g) = Si(OH_4)(g) \tag{2}$$

This silica volatilization causes the recession of the surface of the component, which has been estimated to be as high as ~1 μm/h under normal gas turbine operating conditions [16,17]. Since such components are expected to withstand at least 30,000 h of service without maintenance, this level of corrosion is unacceptable.

Environmental barrier coatings (EBCs) were then developed to negate the pernicious effects that CMAS and steam have on CMCs components [18]. The current generation of EBCs generally presents a rare earth silicate top layer, providing direct protection against CMAS attack and silica volatilization. Several compositions have been suggested and studied [18], each one with its own set of advantages and disadvantages. Among those compositions, ytterbium disilicate (referred to as YbDS in this work) presents several promising characteristics. Its coefficient of thermal expansion is closely matched to that of SiC substrates (3.6 – 4.5 x $10^{-6}$ $K^{-1}$ for YbDS [19] and 4.5 – 5.5 x $10^{-6}$ $K^{-1}$ for SiC [20]), it presents no phase transformation at high temperatures [21], adequate silica volatilization and low thermal conductivity at high temperatures (~2 W·m/K at 1000 °C) [22].

Several studies have reported the interaction between heated steam and YbDS, although currently, there is no standard that allows easy and direct comparison. Additionally, the presence of alumina tubes in most testing rings influences the results through the presence of Al-containing contamination [23–26]. Presentation of the YbDS testing material is varied, ranging from cold pressed pellets [24,27], hot pressed pellets [28–30], coatings formed through oxidation bonded by reaction sintering (OBRS) [31] or coatings deposited using air plasma spraying (APS) [32,33]. Different deposition techniques result in differences in phase composition, microstructure, and porosity levels. Porosity, in particular, presents an interesting dilemma. Traditional EBCs aim for low levels of porosity to prevent steam reaching the substrate through connected pores; however, certain applications require higher levels of porosity. Abradable coatings, for instance, are designed to present porosity as high as 20 % in order to be eroded and allow the turbine blades to create a tight seal without risking damage due to friction [34]. Despite the interest in abradable EBCs, no study has reported the effect that porosity has on the resistance to steam exposure of rare earth silicates.

In this work, three free-standing YbDS coatings deposited using air plasma spraying (APS), with three varying levels of porosity content, were studied. To evaluate the degree of degradation experienced, all three coatings were exposed to a flowing atmosphere of 90 vol.% $H_2O$/10 vol.% $O_2$ with a flow velocity of ~ 100 mm/s, atmospheric pressure and exposure time of 96 h. Two different tests were conducted, at 1350 °C and 1400 °C, to investigate the effect that temperature has on the corrosion from steam with presence of gaseous Al-containing impurities. Temperatures and exposure duration were chosen in line with OEM testing protocols and according to guidance from the high temperature community.

## 2. Experimental methods

### 2.1. Materials and steam exposure

Three different free-standing YbDS coatings were studied in this work, all of them produced using air plasma spraying. The first one, labelled in this work as EBC SG-100 (EBC stands for environmental barrier coating), was a ~350 µm thick YbDS environmental barrier coating, which had a 1 wt.% of alumina powder added to the feedstock prior to spraying. The coating was produced using Treibacher

Industrie AG (Althofen, Austria) YbDS powder through a Praxair Surface Technology (Danbury, Connecticut, USA) SG-100 plasma spray gun. The second free-standing coating, labelled as ABR SG-100 (ABR is short for abradable), was a ~500 µm thick YbDS abradable coating which had a ~1.5 wt.% of polyester powder added to the feedstock prior to spraying. The coating was also deposited using a Praxair Surface Technology SG-100 plasma spray gun, but this time the YbDS powder was provided by Oerlikon Metco AG (Pfäffikon, Freienbach, Switzerland). The third free-standing coating, labelled here as ABR F4, was a ~1000 µm thick YbDS abradable coating which also had a ~1.5 wt.% of polyester powder added to the feedstock prior to spraying. The coating was deposited using an Oerlikon Metco F4 plasma spray gun using the same Oerlikon Metco YbDS powder as with ABR SG-100. Both the Treibacher and Oerlikon YbDS powders were manufactured to conform to a nominal composition of 22 – 24 wt.% $SiO_2$ and balance of $Yb_2O_3$, with a maximum of 5 vol.% of unreacted $Yb_2O_3$, YbMS and $SiO_2$. Compositions were in-line with commercial applications in future gas turbine engines. All three free-standing coatings were cut to produce samples with dimensions of 2.5 x 1.5 $cm^2$. Previous to any steam exposure, all of the as-sprayed samples were heat treated in order to crystallize the amorphous content. Samples were annealed at 1200 °C for 2 h in air, with a heating rate of 5 °C/min.

For the steam exposure, a custom steam rig was designed and built, comprised of the different components shown in Figure 1. The base of the steam rig is an Elite Thermal Systems Ltd (Market Harborough, UK) TSH15/25/180 tube furnace with an alumina tube with an internal diameter of 25 mm. An oxygen bottle is connected to one of the open ends of the tube, being the oxygen flow controlled with a MKS Instruments Inc. (Andover, Massachusetts, USA) type 247 mass flow controller. Deionized water is introduced in the furnace via a Watson-Marlow (Falmouth, UK) 120S peristaltic pump with 0.63 mm bore PVC tubing. The deionized water is first passed through a Grant Instruments (Shepreth, UK) Optima TC120 heated circulating bath kept at 60 °C to facilitate the evaporation once inside the furnace. Both the oxygen and water line are connected through a T connector outside the furnace, being the mixture introduced into the tube furnace using an alumina tube (99.7 % purity) with an internal diameter of 1.5 mm, supplied by Almath Crucibles ltd. (Newmarket, UK), reaching until the start of the hot zone. The other end of the tube furnace was kept open to maintain atmospheric pressure. Oxygen and water flow were set to obtain a flowing atmosphere at the hot zone with a content of 90 vol.% $H_2O$/10 vol.% $O_2$, and a gas velocity of ~100 mm/s. Oxygen and water flow were started once the furnace reached the desired temperature. Samples where placed on top of a alumina plate in the order shown in Figure 1.

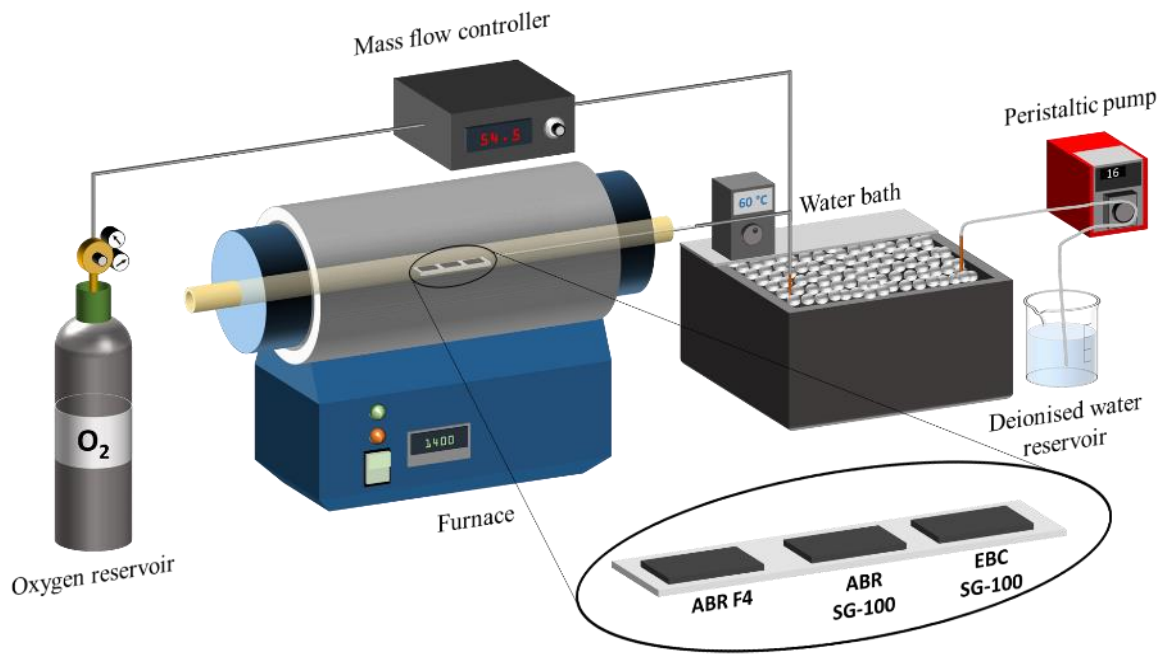

*Figure 1: Schematic of the custom steam rig designed and built for the steam exposure*

Two different exposures were conducted, the first one at 1350 °C and the second one at 1400 °C with a heating rate of 10 °C/min, all the other parameters were kept constant between the two experiments.

2.2. Material characterization

Phase identification of the feedstock powder and free-standing coatings was performed using a Bruker D8 Advance Da Vinci diffractometer (Billerica, Massachusetts, USA) with Cu cathode (wavelength of 1.5406 Å) using Bragg-Brentano geometry. The angular range investigated was from 10° to 70° with a step size set to 0.02° and a dwell time of 0.3 s for all the measurements. Quantitative Rietveld refinement (TOPAS V5, Bruker, Germany) was employed to determine the quantity of each phase [35].

To investigate the microstructure of the free-standing coatings, the samples were cold mounted using Struers EpoFix resin and hardener (Copenhagen, Denmark), then ground and polished to a 1 µm finish using Buehler SiC grinding papers (Leinfelden-Echterdingen, Germany). Scanning electron microscope (SEM) images were taken using a FEI Europe Quanta 600 (Eindhoven, Netherlands). Porosity was calculated as the average measurement across three backscattered electron (BSE) images of representative regions of the coating. All the images were taken with a magnification of 400x, accelerating voltage of 15 kV and spot size of 5 nm. To do the porosity measurement, the open source software "ImageJ" with the image processing package "Fiji" was used [36]. To do so, BSE images were converted into black and white maps upon setting a threshold. Then, the automated function "Analyze particle" was employed, which measured the area percentage of the image covered by porosity, returning an overall value per image. An average of the three images of each coating was calculated, being the standard deviation used as the error.

3. Results

    3.1. Powder and coating characterization

Powder morphology was investigated through SEM imaging to better understand the microstructure of the produced coatings. Figure 2 shows a backscattered image of the two powders used in this study.

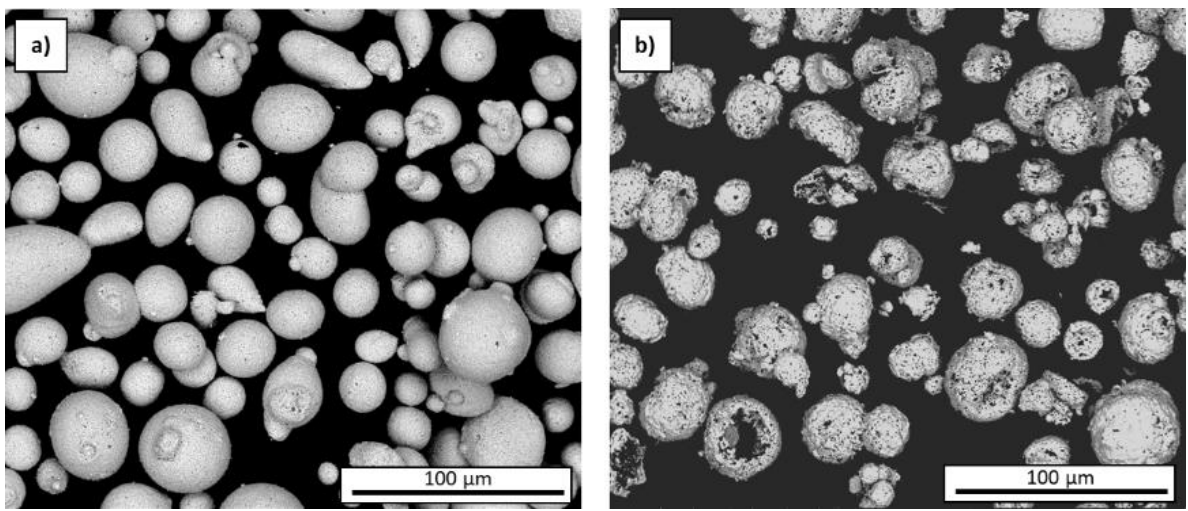

Figure 2: BSE images of the powders used in this work: a) corresponds to sample EBC, b) corresponds to samples ABR

The morphology of the EBC powder, as shown in Figure 2a, is spherical in shape with a smooth surface, mostly lacking any defects or inclusion. On the other hand, ABR powder, as shown in Figure 2b, presents a more irregular shape, still mostly spherical, but with a rough surface and the presence of pores and visible hollow cores.

In order to crystallize the amorphous content of the three as-sprayed coatings, they were annealed; the cross-section of the annealed coatings are shown in Figure 3.

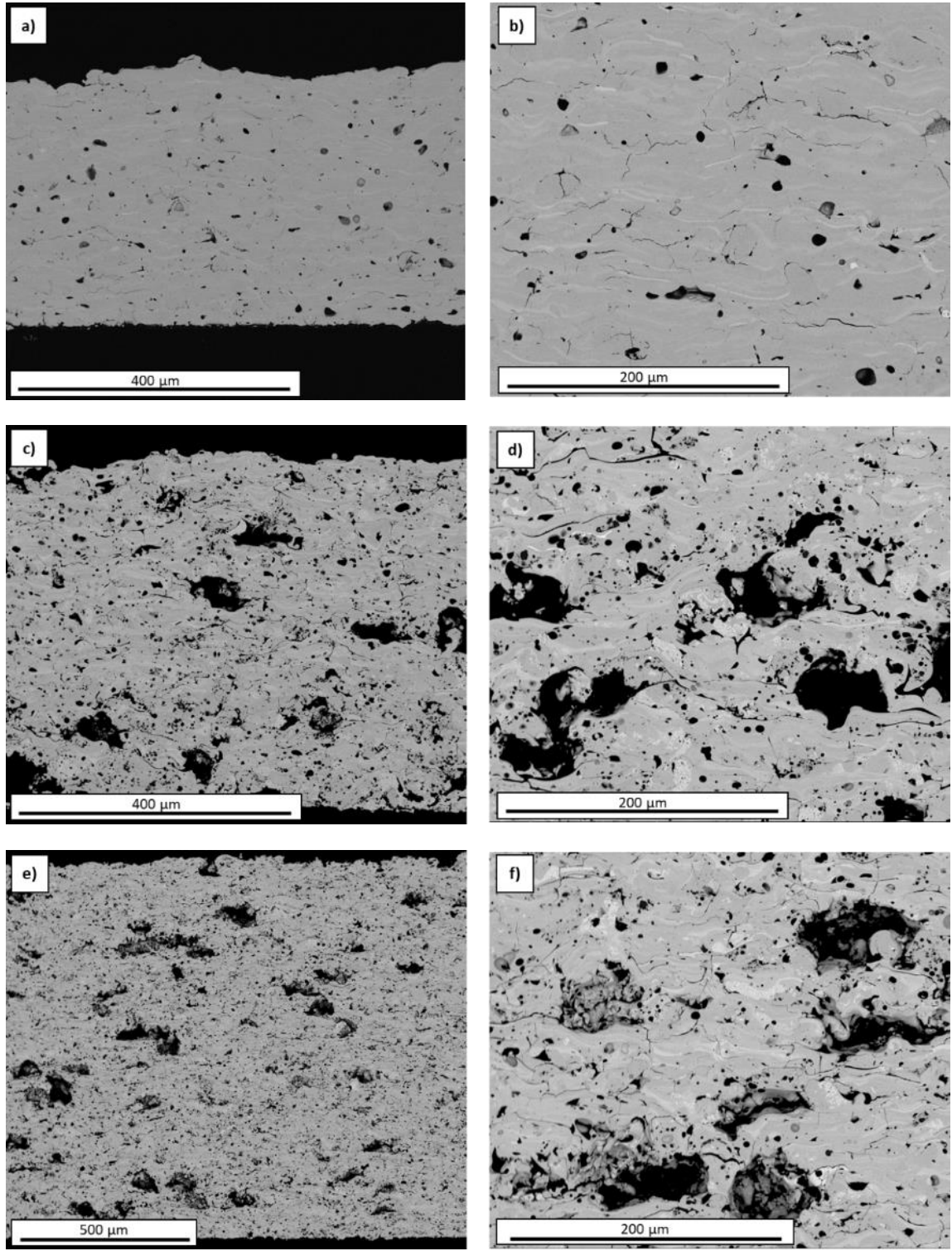

*Figure 3: BSE images of the three free-standing coatings after annealing. Images a) and b) show sample EBC SG-100, c) and d) ABR SG-100 and e) and f) ABR F4*

The cross-section SEM images shown in Figure 3 reveal that sample EBC SG-100 has a lower porosity level and thickness when compared to the two ABR coatings. This is confirmed by the porosity measurements, showing a porosity level of 2.4 ± 0.3 % for EBC SG-100. On the other hand, sample ABR SG-100 has a porosity of 21.3 ± 1.1 %, and ABR F4 has 19.4 ± 4.0 %. The higher porosity in the two ABR coatings is a direct consequence of the addition of polyester in the feedstock powder. During the annealing, the polyester burns off, leaving empty pores behind. As for the thickness, measurements of sample EBC SG-100 revealed a thickness of 368.6 ± 10.4 µm, while ABR SG-100 was 509.9 ± 9.2 µm and ABR F4 was 1099.7 ± 12.7 µm. The different thicknesses were chosen to represent the typical values required for each application in service, with EBCs traditionally remaining below 400 µm, whereas abradable coatings are generally thicker.

The top surface of the annealed samples were investigated in an SEM to examine the morphology and provide a baseline to which exposed samples could be compared. The SEM images can be seen in Figure 4.

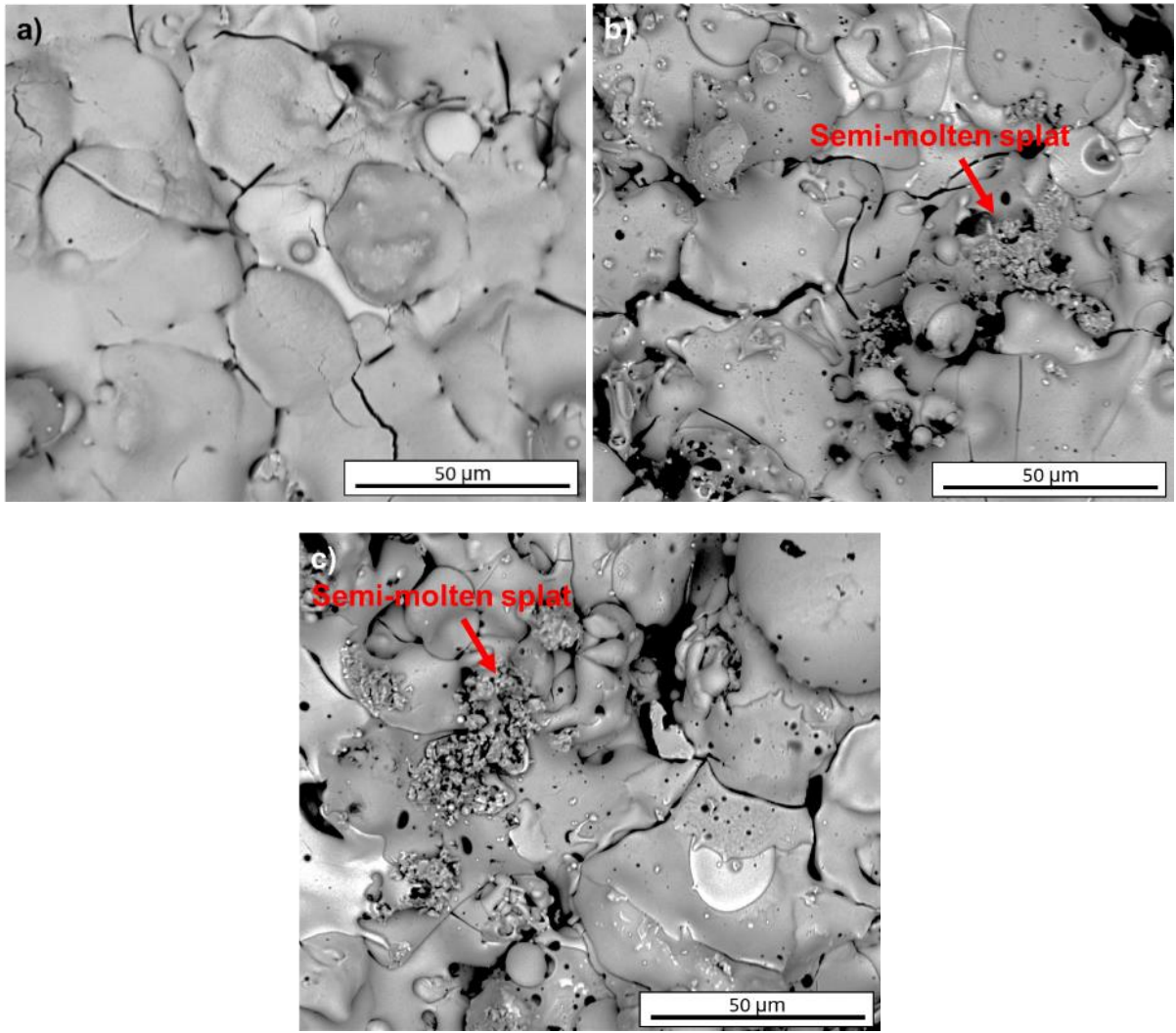

*Figure 4: BSE images of the top surface of the annealed samples. Image a) corresponds to sample EBC SG-100, image b) to ABR SG-100 and image c) to ABR F4*

From the images of the top surface shown in Figure 4, it can be seen that sample EBC SG-100 presents a smooth surface with well-molten splats, along with cracks distributed across the surface. In the case of the two abradable samples, ABR SG-100 and ABR F4, Figure 4b and c respectively, the surface presents a combination of well molten and semi-molten splats, the latter being labelled in the images.

The phase composition of the annealed coatings was also studied. The XRD diffractogram for the feedstock powder, as-sprayed and annealed conditions are shown in Figure 5 for each of the three coatings.

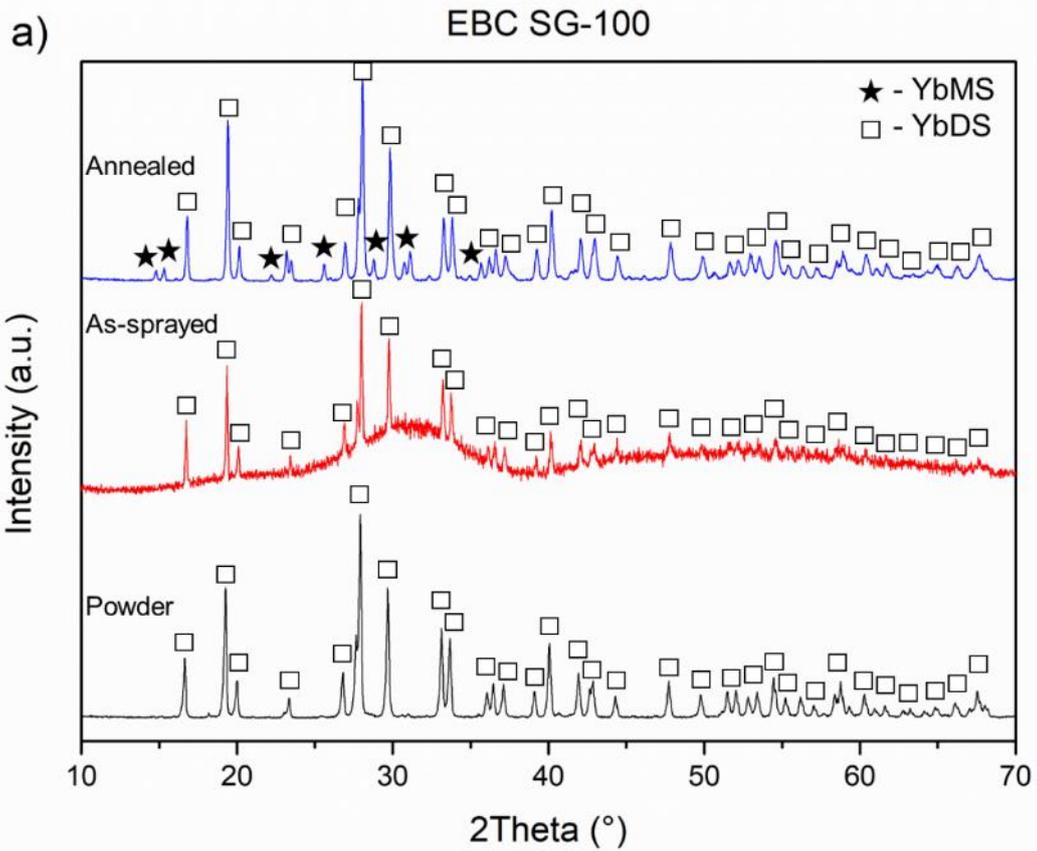
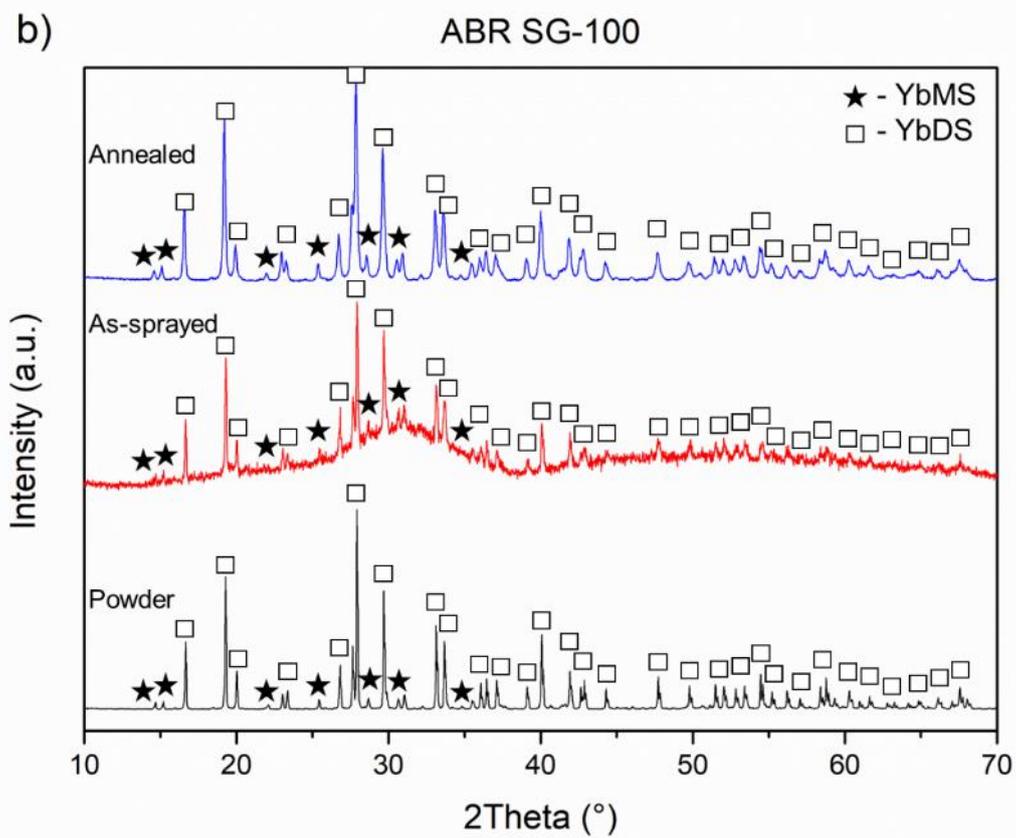

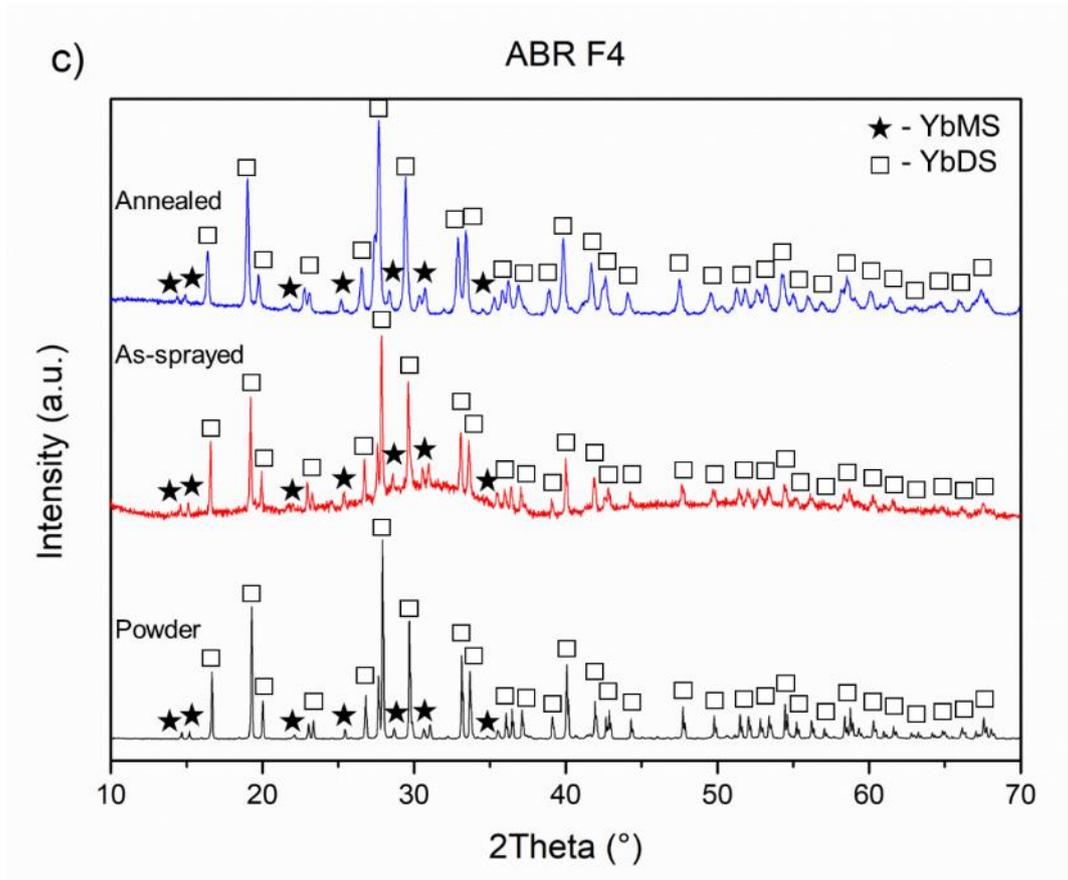

*Figure 5: XRD measurements for a) EBC SG-100, b) ABR SG-100 and c) ABR F4. On each graph, the bottom plot corresponds to the powder, middle to the as-sprayed coating and top to the annealed coating. Phases have been identified with a star (★) for YbMS and a square (□) for YbDS*

The XRD measurements show that the EBC powder, Figure 5a, presents only peaks from the YbDS phase (PDF card number 00-082-0734), although Rietveld refinement identified a small quantity of YbMS (5.4 wt.%). The powder used to spray the abradable samples, Figure 5b and c, shows the presence of a higher quantity of YbMS (PDF card number 00-040-0386), measured as 18.7 wt.%. In the as-sprayed condition, EBC SG-100 only shows the presence of YbDS peaks with a 63.6 wt.% of amorphous content, whereas samples ABR SG-100 and ABR F4 show YbMS peaks accounting for ~25 wt.%. In addition to the crystalline peaks, all of the three as-sprayed coatings show two broad amorphous humps, centered on ~30° and ~55°. The annealing process removes the presence of these amorphous humps, showing all of the three samples consist mainly of YbDS, with approximately ~30 wt.% of YbMS. The detailed phase content as measured through Rietveld refinement is presented in Figure 11.

## 3.2. Steam exposure at 1350 °C

The cross-section of the samples exposed to steam at 1350 °C for 96 h is shown in Figure 6.

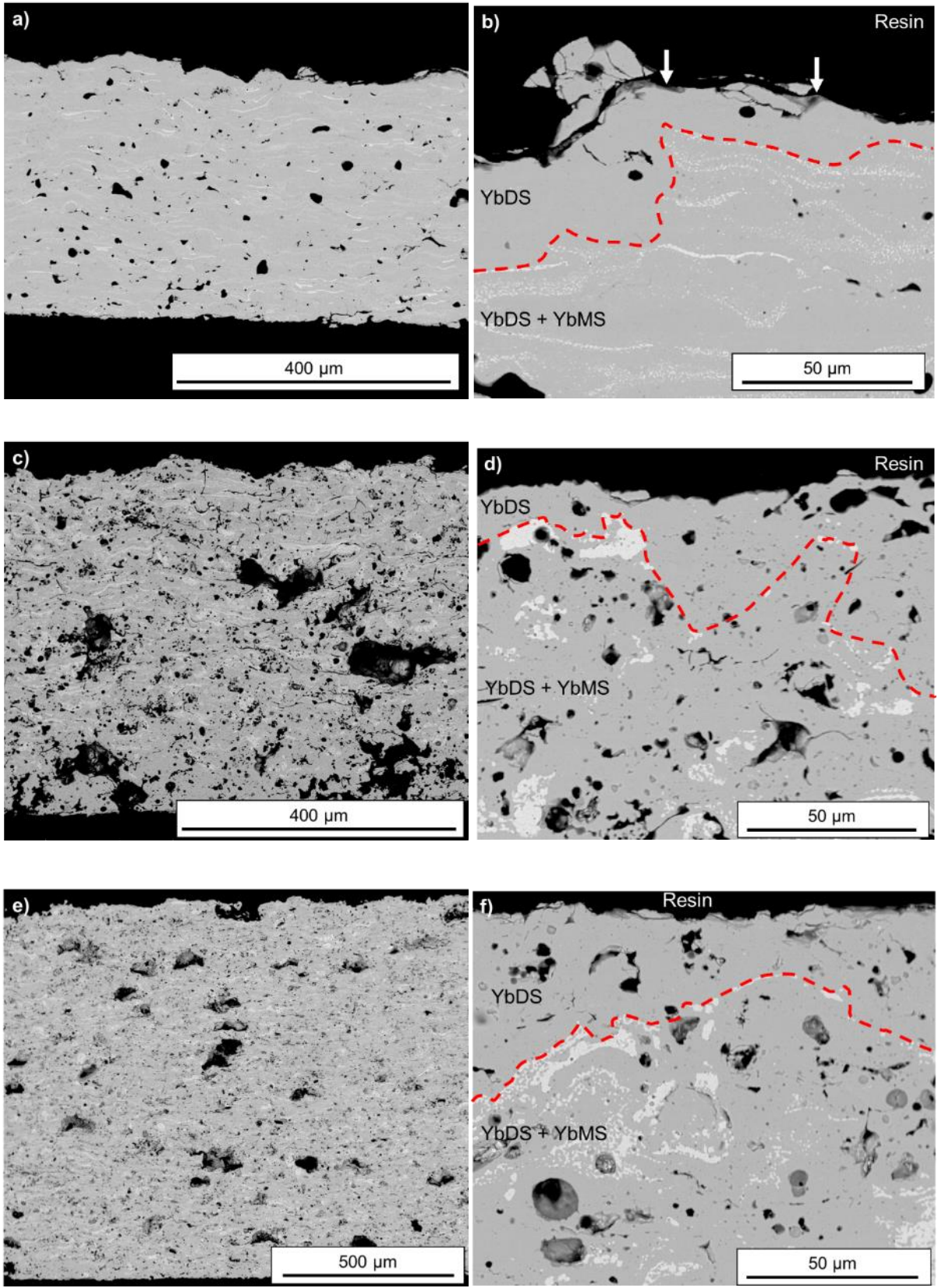

*Figure 6: Low and high magnification BSE images of the cross-section of the samples exposed to steam for 96 h at 1350 °C. Images a) and b) correspond to sample EBC SG-100, images c) and d) to ABR SG-100 and images e) and f) to ABR F4. The red dashed line marks where no more YbMS could be found. White arrows in image b) indicate where a new phase was detected*

From the SEM images shown in Figure 6, it can be seen that exposure to steam at 1350 °C caused the depletion of the YbMS closest to the surface, leaving behind a YbMS depleted layer. In the case of sample EBC SG-100, the high magnification image in Figure 6b shows small amounts of a new phase (marked with arrows) present at the surface of the coating. Due to the size of these features and the overlap between the Al-$K_\alpha$ and Yb-$M_\alpha$ lines, EDS quantification proved to be challenging. More details and identification are provided in the section 3.3 "Steam exposure at 1400 °C".

In addition to the cross-section, the top surface of the exposed samples was also imaged. The images are shown in Figure 7.

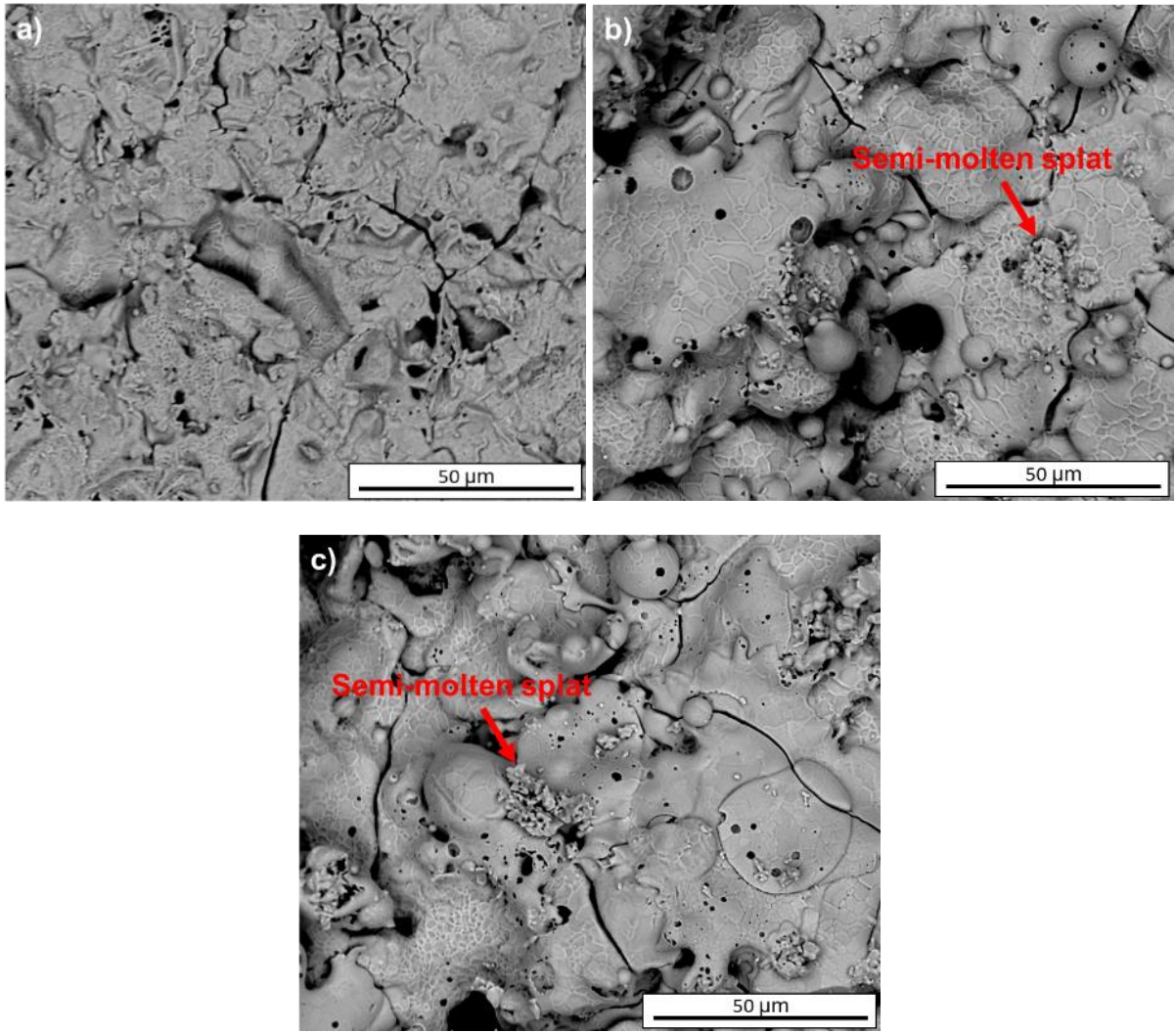

*Figure 7: BSE images of the top surface of the samples exposed to steam at 1350 °C for 96 h. Image a) corresponds to sample EBC SG-100, image b) to ABR SG-100 and image c) to ABR F4*

From the images of the top surface shown in Figure 7, it can be seen that sample EBC SG-100, Figure 7a, has a smoother aspect, meaning the individual splats are more difficult to distinguish. On the other hand, samples ABR SG-100 and ABR F4, Figure 7b and c, still show the presence of individual splats, both molten and semi-molten. All of the three samples show the effects of steam exposure; the grain boundary attack on the splats is clearly visible. Multiple cracks, both intra- and inter-splat, are also visible on the three samples.

### 3.3. Steam exposure at 1400 °C

SEM images of the cross-section of the three coatings exposed to steam at 1400 °C for 96 h are presented in Figure 8.

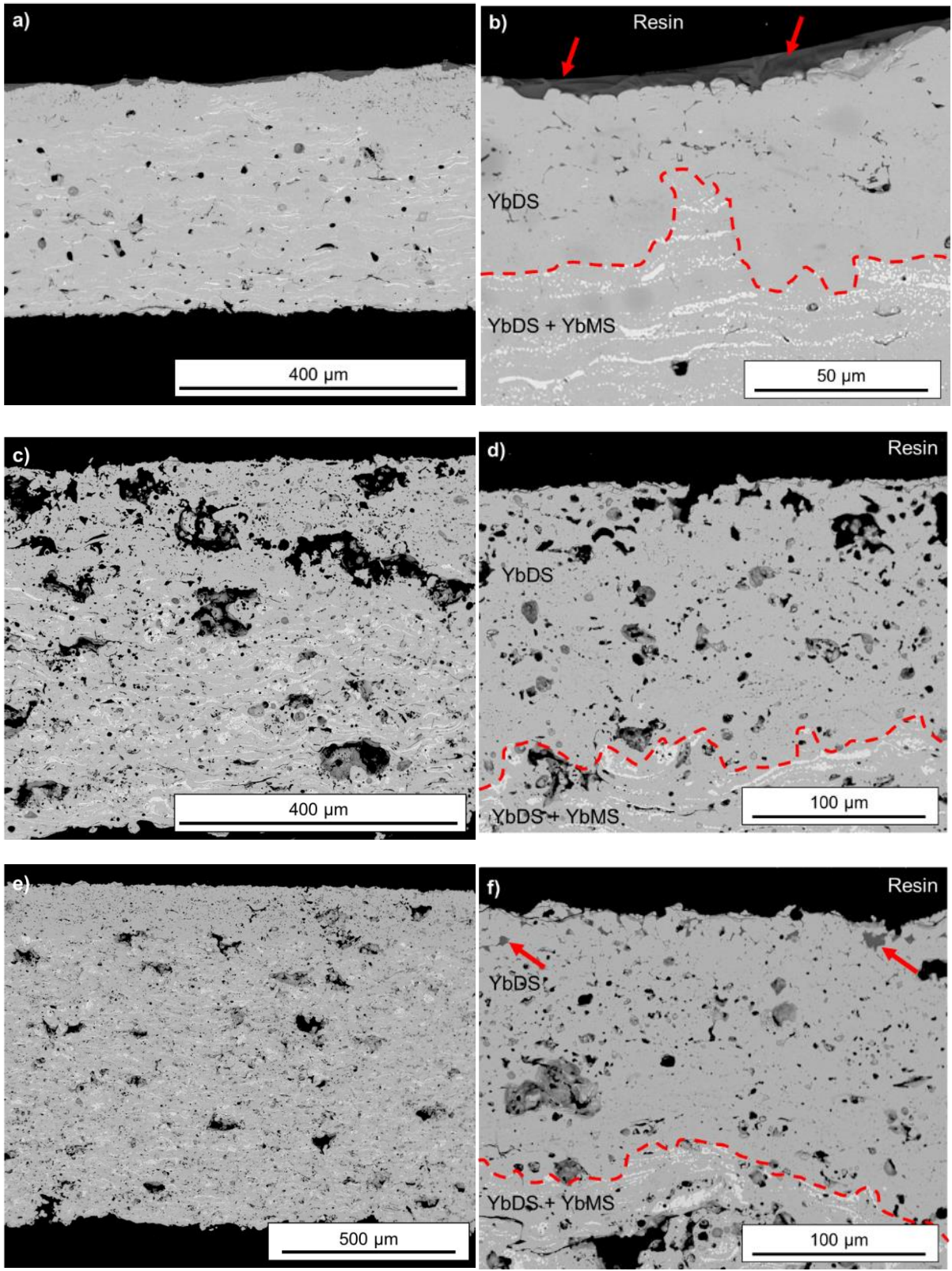

*Figure 8: Low and high magnification BSE images of the cross-section of the samples exposed to steam for 96 h at 1400 °C. Images a) and b) correspond to sample EBC SG-100, images c) and d) to ABR SG-100 and images e) and f) to ABR F4. The red dashed line marks where no more YbMS could be found. Red arrows in images b) and f) indicate where a new phase was detected*

From the low and high magnification images of the three coatings exposed to steam at 1400 °C, as shown in Figure 8, some differences can be appreciated. First, there is a clear new phase on the surface of sample EBC SG-100, marked with red arrows in Figure 8b. This scale, visible even at low magnification, was only present to such an extent on this sample. Smaller traces could be found within the coatings, especially in ABR F4, as marked with red arrows in Figure 8f. In the case of the abradable samples, this new phase was not seen as a homogenous scale on top of the surface but as small patches and within filled pores. EDS identification was tried (not shown here), but the results were unreliable due to the already mentioned issue of overlapping peaks between the Al-$K_\alpha$ and Yb-$M_\alpha$ lines, which is why XRD measurements were used for identification, as shown in Figure 10.

Another feature observed was the previously mentioned YbMS depleted layer near the surface of the coating. The thickness of the depleted layer is thinner in the case of sample EBC SG-100, approximately 50 µm, whereas the thickness of the depleted layer on the abradable samples is closer to ~125 µm. In addition, the top area of the depleted layer presents visible inter-splat boundaries, not present in the samples exposed to steam at 1350 °C, as it can be seen in Figure 6. This newly formed feature were located at the inter-splat boundaries and can be more clearly seen in sample EBC SG-100 due to the initial low level of porosity. The top surface of the three coatings exposed to steam at 1400 °C was imaged, as shown in Figure 9.

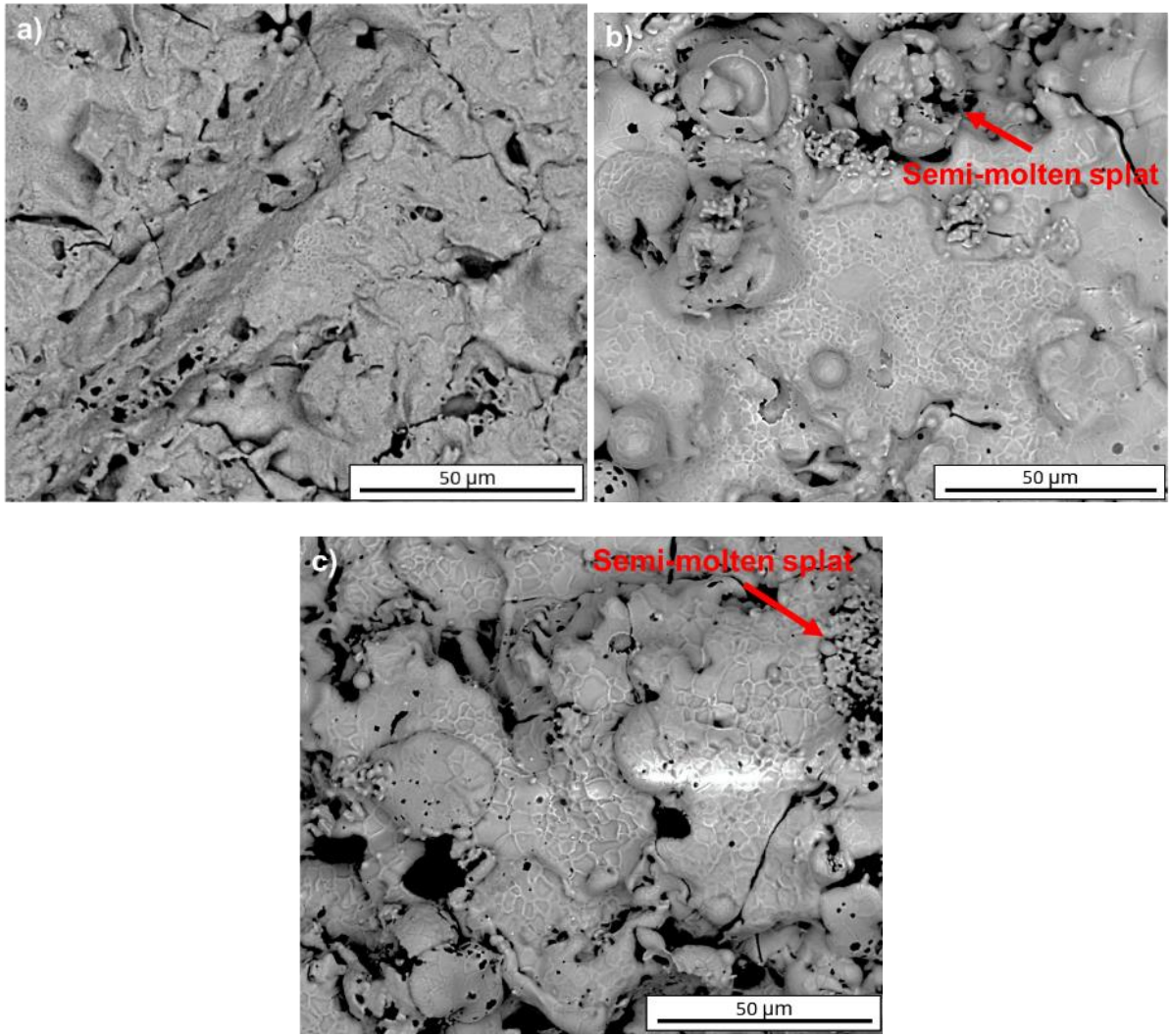

*Figure 9: BSE images of the top surface of the samples exposed to steam at 1400 °C for 96 h. Image a) corresponds to sample EBC SG-100, image b) to ABR SG-100 and image c) to ABR F4*

From the SEM images of the top surface, the grain boundary attack is less visible in the EBC SG-100 sample exposed to 1400 °C (Figure 9a) compared to 1350 °C (Figure 7a). Individual splats are difficult to identify, whereas cracks are visible. The top surface of the two abradable coatings, ABR SG-100 (Figure 9b) and ABR F4 (Figure 9c), show very similar features, with presence of semi-molten splats. Individual splats affected by grain boundary corrosion can be seen, as well as cracks.

XRD measurements were performed to study the phase content of the three coatings. Figure 10 shows the XRD measurements of the three coatings on the three different conditions studied: before exposure, after exposure at 1350 °C and after exposure after 1400 °C.

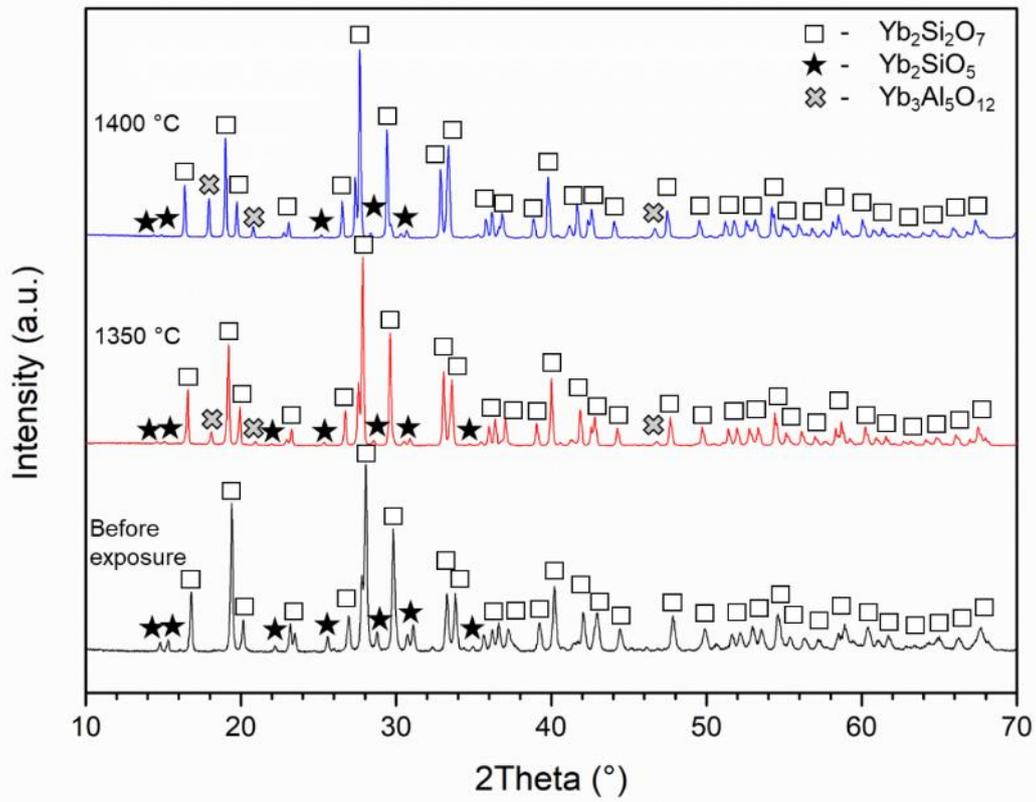

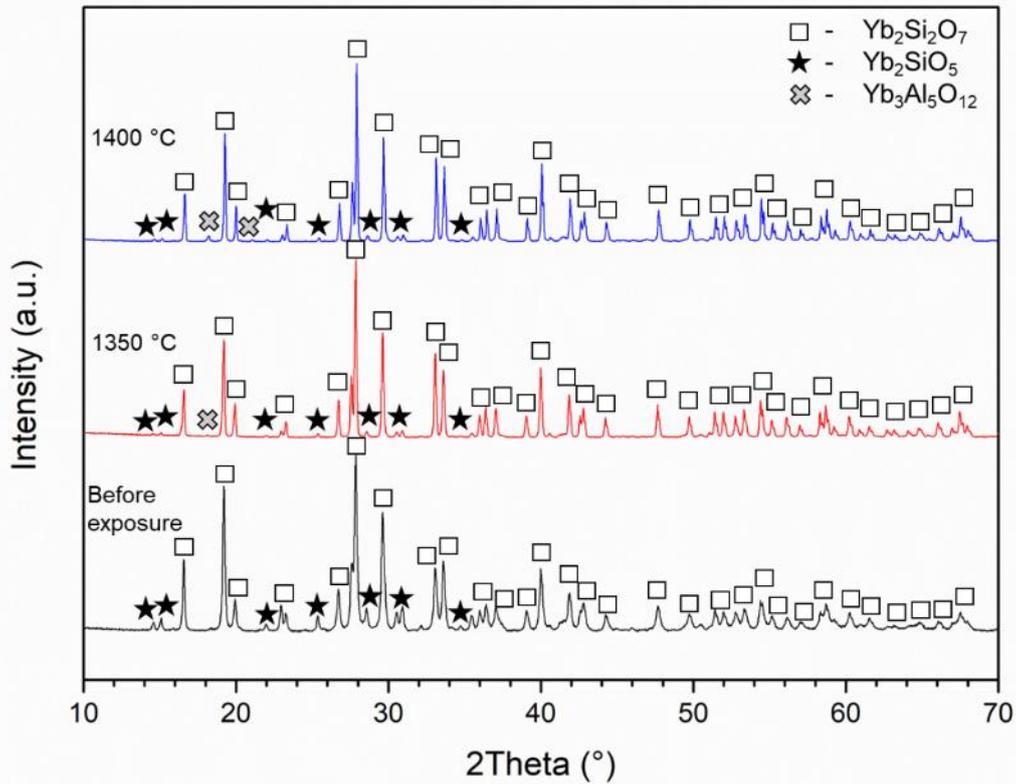

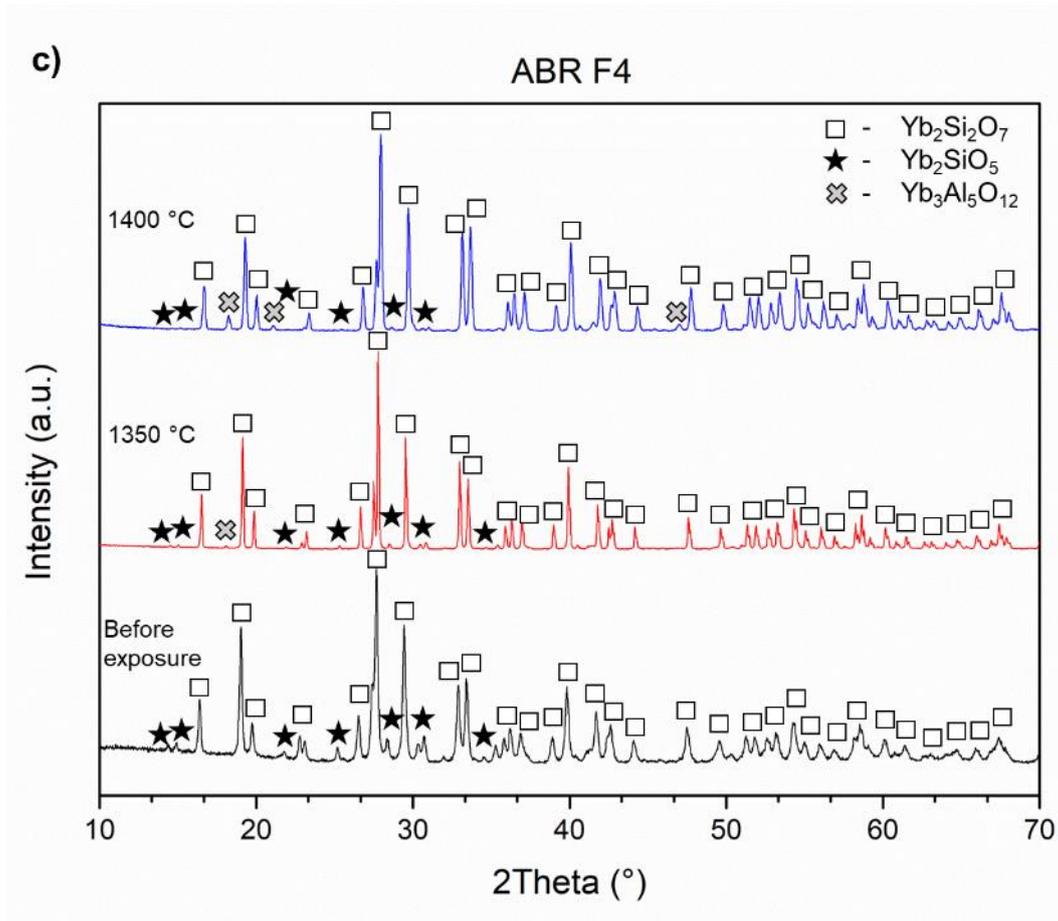

*Figure 10: XRD measurements for a) EBC SG-100, b) ABR SG-100 and c) ABR F4. On each graph, bottom plot corresponds to the coating before steam exposure, middle to the 1350 °C steam exposure and top to the 1400 °C steam exposure. Phases have been identified with a square (□) for YbDS, a star (★) for YbMS and a cross (✖) for the garnet*

The XRD measurements for sample EBC SG-100, Figure 10a, show how once the sample is exposed to steam at 1350 °C, there is a reduction in the intensity of the YbMS peaks. For EBC SG-100, this means a reduction from a YbMS content of 29.3 wt.% down to 8.1%, as shown in Figure 11. In the case of the two abradable samples, ABR SG-100 and ABR F4, the reduction is from ~29 wt.% to ~12 wt.%. At the same time, a new phase can be seen, identified as ytterbium garnet ($Yb_3Al_5O_{12}$, PDF card number 00-023-1476). This new phase was seen on the SEM images of the cross-section, Figure 6 and Figure 8. The presence of garnet is most predominant on EBC SG-100, with a 7.0 wt.%, whereas the two abradable samples present a garnet phase content around or below 1 wt.%. This trend continues for the samples exposed at 1400 °C. EBC SG-100 experiences a further reduction in the YbMS content, down to 3.9 wt.%, with garnet content rising to 17.6 wt.%. Regarding ABR SG-100, exposure to steam at 1400 °C did increase the amount of garnet formed (2.6 wt.%) compared to exposure at 1350 °C (0.1 wt.%). This change is most notable in ABR F4, having a garnet content of 1.0 wt.% when exposed at 1350 °C, rising to 7.2 wt.% when exposed at 1400 °C. It should be noted that the phase content only takes into account the top 10 -15 µm of the coatings due to the estimated penetration depth of x-rays for this composition [23].

The phase content as quantified through Rietveld refinement for all the samples here studied can be found in Figure 11.

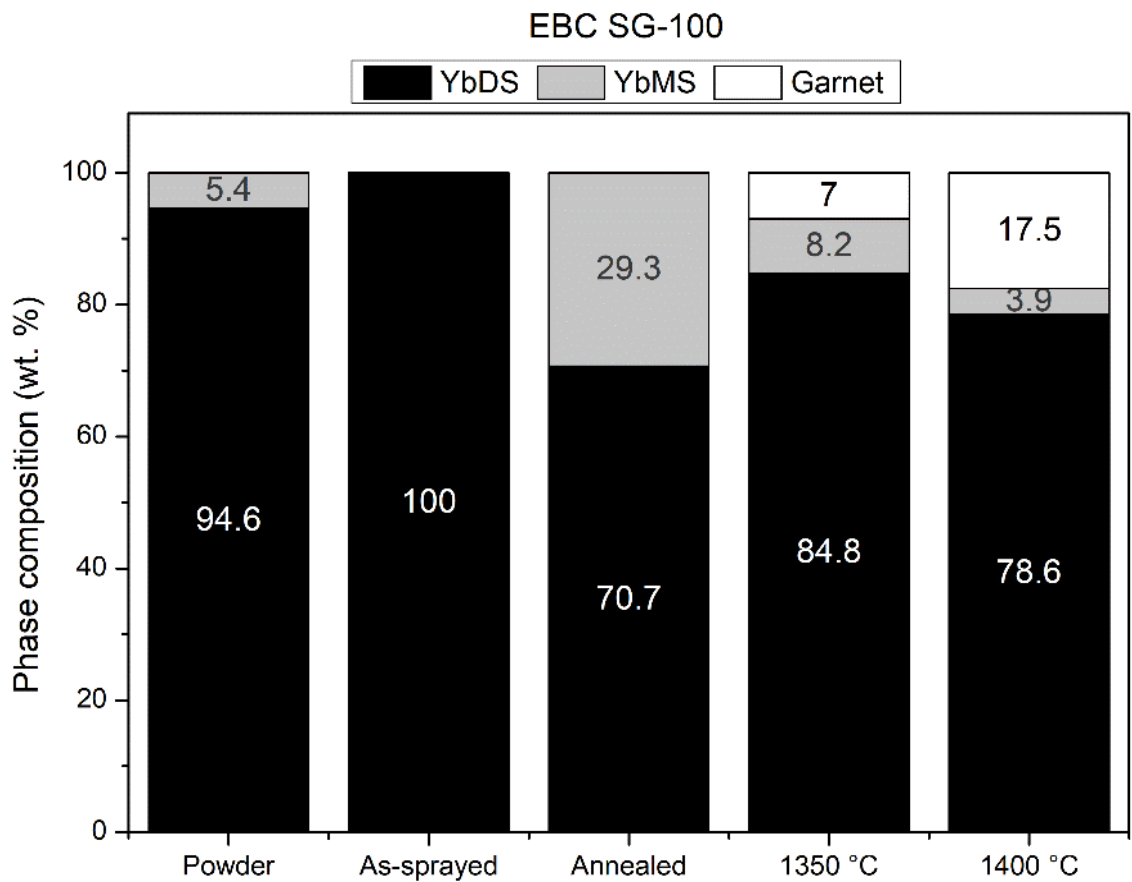

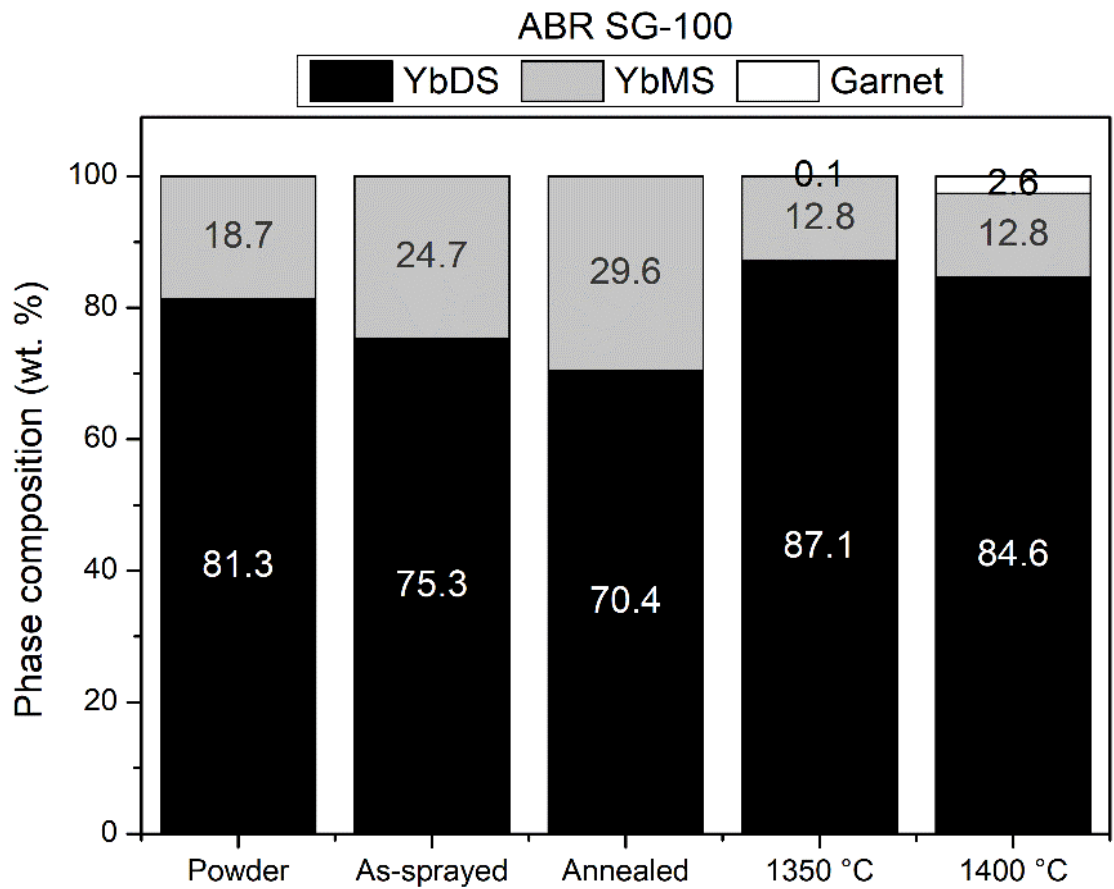
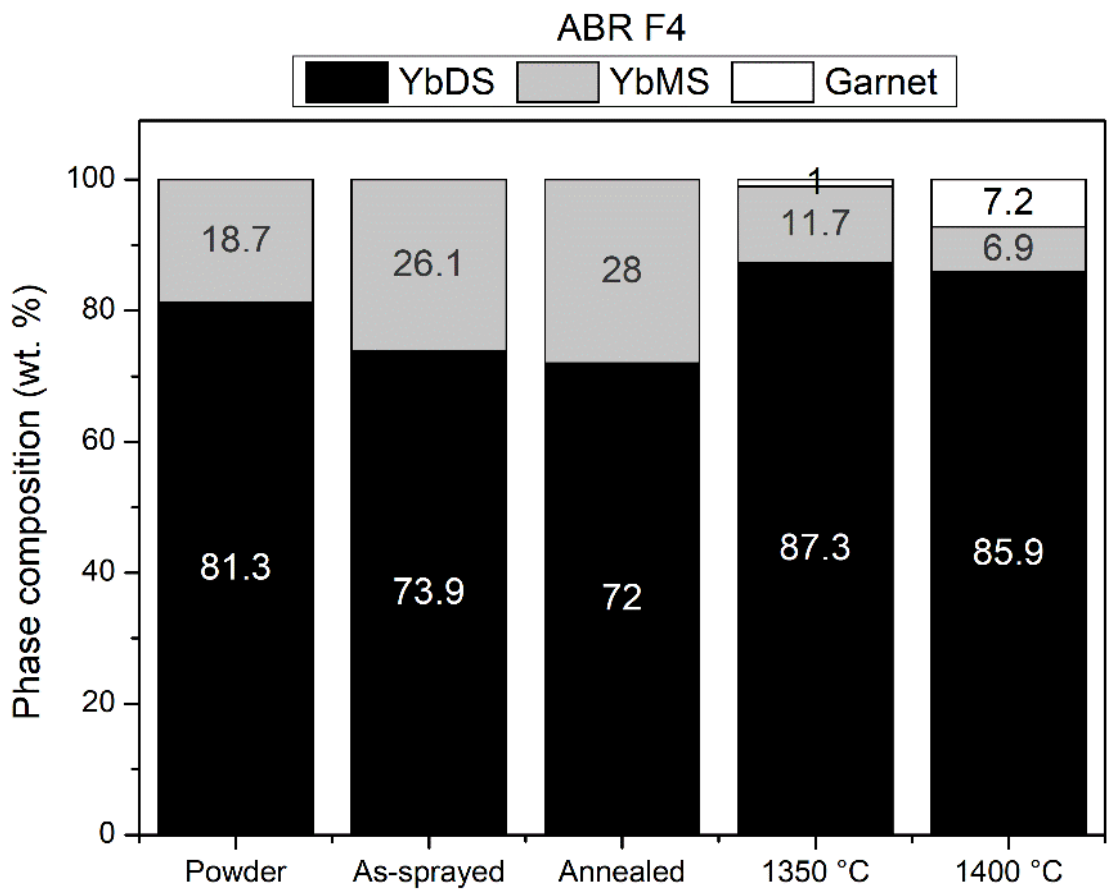

*Figure 11: Crystalline phase content quantified using Rietveld refinement for coatings EBC SG-100, ABR SG-100 and ABR F4*

## 4. Discussion

Steam degradation is correlated with several factors, such as coating phase composition, microstructure, temperature and steam velocity [18]. The three coatings studied here present a similar composition, as revealed by the XRD measurements in Figure 5. EBC powder was mainly composed of YbDS, whereas ABR powder had traces of YbMS phase. During APS deposition, preferential volatilization of $SiO_2$ takes place due to the in-flight conditions, leading to the formation of small quantities of YbMS [37]. This phase is in an amorphous state due to the rapid cooling experienced by the splats upon impact [38]. This would explain why no distinguishable YbMS peaks can be found on the as-sprayed EBC SG-100 sample (Figure 5a), where the amorphous content is as high as 63.6 %. Once the annealing treatment is completed, the amorphous content is crystallized, leading to only crystalline peaks for YbDS as the main phase and small quantities of YbMS.

Despite this very similar starting composition of the annealed coatings, microstructure presents a differentiating factor between the EBC SG-100 sample and the two ABR coatings. As shown in the cross-section SEM images in Figure 3, EBC SG-100 presented a lower porosity level (2.4 ± 0.3 %) which can be explained by the absence of pores and defects in the feedstock powder as well as the lack of added polyester as pore former, as seen in Figure 2a. On the other hand, both the abradable coatings, ABR SG-100 and ABR F4, present much higher levels of porosity, 21.3 ± 1.1 % and 19.4 ± 4.0 %, respectively. This is due to the presence of porosity and hollow cores in the feedstock powder (Figure 2b) and the addition of polyester as a pore former. When comparing the top surface of the annealed samples, there is also a clear difference between the EBC samples and the abradable ones, as can be seen in Figure 4. EBC SG-100 presents a smoother surface caused by well-molten splats that flattened upon impact. The abradable samples show the presence of semi-molten, but not completely flat splats, giving rise to a rougher surface where individual splats are easily identifiable. In all of the three samples, intra- and inter-splat cracks could be found.

From the SEM images of the top surface of the three steam exposed coatings, shown in Figure 7, it can be seen that the interaction between the steam (including the presence of gaseous Al-containing impurities) and the coatings takes place preferentially at the grain boundaries, a phenomenon also reported by Maier *et al.* [24] and Rohbeck *et al.* [25]. Exposure to steam at 1350 °C caused the appearance of a YbMS depleted layer on all of the three coatings, as shown in Figure 6, being particularly visible in the high magnification images. In addition to the depletion of YbMS from the top layers near the surface, XRD measurements indicate the formation of a new phase, identified as ytterbium garnet. Figure 11 shows the evolution of the quantitative phase content for each of the three coatings here studied. In all three coatings, a reduction in the YbMS content could be observed as the content of garnet increased.

Exposure of YbDS to flowing, high temperature steam has been reported to cause $SiO_2$ volatilization and YbMS formation through the reaction shown in Equation 3 [26,32,33,39–41].

$$Yb_2Si_2O_7 + 2H_2O(g) \rightarrow Yb_2SiO_5 + Si(OH)_4(g) \quad (3)$$

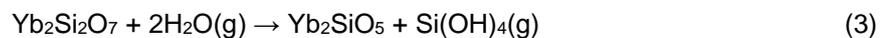

Nevertheless, in this work, no presence of a YbMS layer could be observed. Instead, a reduction in the YbMS content, including a YbMS depleted layer at the surface of the coatings, along with the appearance of garnet is detected. The appearance of garnet on steam exposure testing conducted using high purity alumina tubes is expected and has been extensively reported in the literature [23–25]. In particular, Kane *et al.* [23] report the formation of a YbMS depleted layer on a multi-layer YbDS/Si EBC exposed to steam at 1300 °C. Their system, containing a Si bond-coat, presented YbMS depletion both near the surface and at the Si – YbDS interface, suggesting the involvement of Si/$SiO_2$ in the YbMS depletion process, but they reported no presence of garnet at this temperature. As can be seen in Figure 10, XRD measurements confirmed the presence of garnet at 1350 °C in all of the three samples, although sample EBC SG-100 showed the highest content. This seems to suggest two things: first, the formation of garnet is temperature dependent, with 1300 °C not high enough for the reaction to take place. Secondly, the 1 wt.% of alumina added to the feedstock EBC powder could explain why this sample shows the highest content of garnet at 1350 °C. Whereas the abradable samples rely on the Al-containing impurities from the furnace tubes to form the garnet via gas phase transport, the EBC SG-100 sample has the additional alumina within the coating.

Regarding the consumption of the YbMS phase to form garnet, similar results have been reported by Kane *et al.* [23] and Rohbeck *et al.* [25]. Although Rohbeck *et al.* did not specify the steam velocity used in their experimental setup, Kane *et al.* measured their steam velocity to be 1.5 cm/s. It is suggested that steam velocity plays a key role in the corrosion mechanism observed [26], with low-velocity flowing steam not causing YbDS volatilization.

The mechanism behind the depletion of YbMS is not fully understood yet, although two mechanisms have been proposed in the literature. The first one involves the consumption of the YbMS as it reacts with the Al-containing impurities (or the alumina present within the coating in the case of sample EBC SG-100) to form the garnet phase. Kane *et al.* [23] suggested that YbMS is more reactive than YbDS regarding alumina, which would explain why YbDS is unaffected in the steam exposure. This mechanism is temperature dependent, as the lowest eutectic point of the Yb-Al-Si-O system is 1500 °C [42], although the presence of alkali impurities from the furnace may unlock the formation of aluminosilicate compounds at a lower temperature [23]. This temperature dependence would explain the increase in YbMS depletion in all the samples in the 1400 °C steam exposure. When considering the 1400 °C exposure of sample EBC SG-100, it is worth noting that this sample was the one where the splat boundaries where more clearly visible, as presented in Figure 8b. For the two abradable coatings, this feature in the depleted layer was more difficult to detect due to the initial higher level of porosity. Similar to the infiltration of CMAS into YbDS [43–46], splat boundaries seem to be the preferential path for the ingress of Al-containing impurities, which would explain this phenomenon. Therefore, it can be suggested that this mechanism is prevalent in the case of sample EBC SG-100, where the extensive formation of garnet is observed. Once a dense scale of garnet is formed on the surface of the coating at 1400 °C, the ingress of Al-containing impurities is hindered, which slowed the expansion of the YbMS depleted layer compared to the abradable samples.

The second mechanism for the depletion of YbMS is based on the reaction between YbMS and $SiO_2$ to form YbDS. The presence of two YbMS depleted layers in work by Kane *et al.* [23], one near the surface and one near the Si bond coat after steam exposure at 1300 °C, supports the idea of a low temperature Si/$SiO_2$ mediated mechanism for the depletion of YbMS. The presence of $SiO_2$ could not be confirmed through XRD measurements, which could indicate that the quantity present is below the detection limit of the technique. The high content of ytterbium on the coatings limits the estimated x-ray penetration to below 10 - 15 µm [23], limiting the amount of $SiO_2$ available for detection. Additionally, the overlap between the YbMS and $SiO_2$ peaks might have masked the presence of traces amounts of $SiO_2$. Nevertheless, since the system here studied was a free-standing coating without the presence of a Si bond coat, and presence of $SiO_2$ could not be observed, this mechanism cannot be robustly argued to explain the phenomenon detected here. A schematic representing the interaction between high-temperature steam and both EBC YbDS coating and abradable coatings is presented in Figure 12.

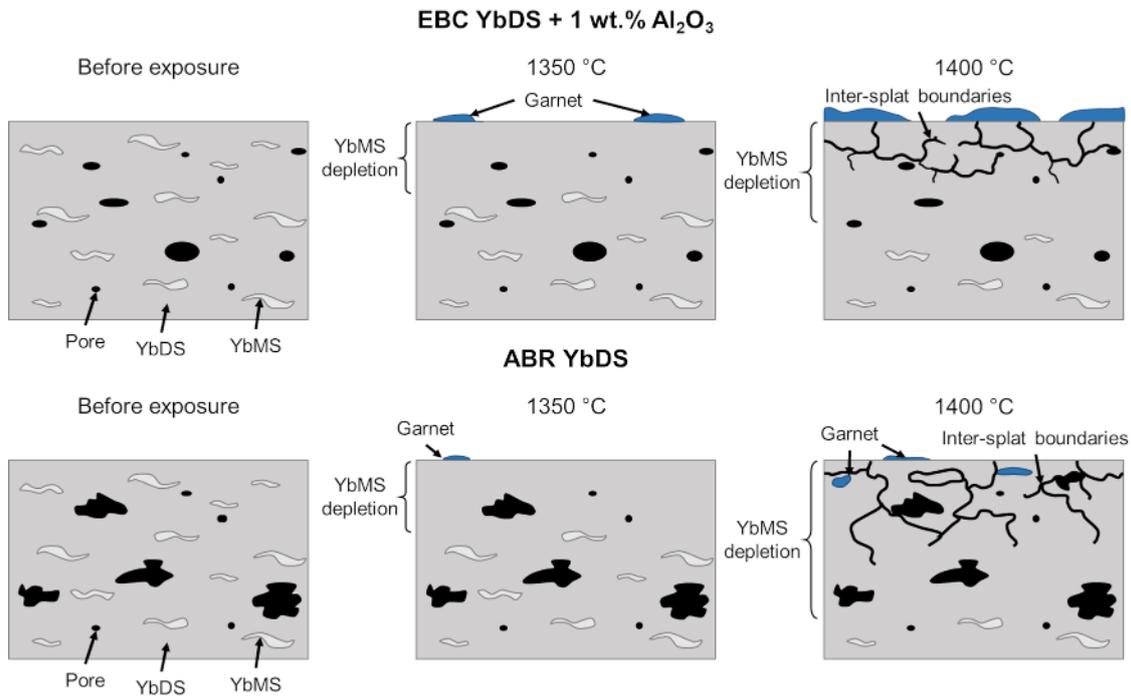

*Figure 12: Schematic with the different behavior of EBC YbDS + 1 wt.% $Al_2O_3$ and abradable YbDS coatings when exposed to steam at high temperature*

From Figure 12, it can be seen how EBC SG-100 presents depletion of YbMS once exposed to steam at 1350 °C, while forming small quantities of garnet located at the top surface. At 1400 °C, the formation of garnet is more extensive, leading to a scale at the surface and a slightly larger depleted layer. Inter-splat porosity can be observed in the depleted layer. In the case of the abradable samples, exposure to steam at 1350 °C also produces a YbMS depleted layer, although lower quantities of garnet are formed due to the absence of added alumina to the coatings. At 1400 °C, however, evidence of garnet formation can be observed, both at the surface and inside pores located near the surface. The depletion layer is considerably larger, containing finer pores aligned with the splat boundaries, along with larger pores. As previously mentioned, the presence of these fine pores within the inter-splat boundaries indicates that splat boundary is the preferential ingress path of gaseous Al-containing impurities.

5. **Conclusions**

Degradation of ytterbium disilicate EBCs under steam conditions is one of the main considerations for the successful implementation of SiC CMC components into the current generation of gas turbine engines. In this work, three free-standing YbDS coatings deposited using APS were exposed to steam at 1350 °C and 1400 °C for 96 h. The results show that sample EBC SG-100, with a low porosity level and 1 wt.% of $Al_2O_3$ added to the feedstock powder, presented moderate depletion of YbMS near the surface with the formation of ytterbium garnet. At 1400 °C, the garnet formed a dense scale at the surface and inter-splat porosity formed within the YbMS depleted layer. The mechanism for the YbMS depletion is believed to be a reaction with gaseous Al-containing impurities from the alumina furnace tubes and the alumina present within the coating, leading to the formation of the garnet.

The two abradable samples, with 1.5 wt.% polyester added to the feedstock powder as a pore former, behaved in a similar fashion at 1350 °C, with the appearance of a YbMS depletion layer and traces amount of garnet detected. At 1400 °C, the size of the depleted layer grew considerably larger than in the case of the EBC sample, with less garnet phase forming. Porosity also increased within the depleted layer; however, its effect was more difficult to clearly identify due to the higher level of porosity. The mechanism for the formation of the YbMS depleted layer is associated with the gaseous Al-containing impurities from the furnace. Since the abradable coatings did not contain added alumina, less garnet phase formed compared to the EBC coating.

In both cases, the exposure to flowing steam for 96 h at high temperature did not produce any evidence that the integrity of the coatings might be compromised, both in terms of cracking or loss of mass. Nevertheless, Al-containing impurities coming from the furnace tubes played a key role in the steam degradation of the coatings, requiring further investigation using experimental setups where external contributions are not a factor.

**Acknowledgements**

This work was supported by the Engineering and Physical Sciences Research Council (EPSRC) (grant number EP/L016206/1). The authors would like to thank Dr Hannah Constantin for her assistance in performing the XRD measurements and to the Nanoscale and Microscale Research Centre (nmRC) at the University of Nottingham for providing access to microscopy facilities.

**References**

[1] N.P. Padture, Advanced structural ceramics in aerospace propulsion, Nat. Mater. 15 (2016) 804–809. https://doi.org/10.1038/nmat4687.

[2] M. van Roode, Ceramic Gas Turbine Development: Need for a 10 Year Plan, J. Eng. Gas Turbines Power. 132 (2010) 1–8. https://doi.org/10.1115/1.3124669.

[3] J. Steibel, Ceramic matrix composites taking flight at GE Aviation, Am. Ceram. Soc. Bull. 98 (2019) 30–33.

[4] K.M. Grant, S. Krämer, G.G.E. Seward, C.G. Levi, Calcium-Magnesium Alumino-Silicate Interaction with Yttrium Monosilicate Environmental Barrier Coatings, J. Am. Ceram. Soc. 93 (2010) 3504–3511. https://doi.org/10.1111/j.1551-2916.2010.03916.x.

[5] K.M. Grant, S. Krämer, J.P.A. Löfvander, C.G. Levi, CMAS degradation of environmental barrier coatings, Surf. Coatings Technol. 202 (2007) 653–657. https://doi.org/10.1016/j.surfcoat.2007.06.045.

[6] D.L. Poerschke, D.D. Hass, S. Eustis, G.G.E. Seward, J.S. Van Sluytman, C.G. Levi, Stability and CMAS Resistance of Ytterbium-Silicate/Hafnate EBCs/TBC for SiC Composites, J. Am. Ceram. Soc. 98 (2015) 278–286. https://doi.org/10.1111/jace.13262.

[7] N.S. Jacobson, J.L. Smialek, D.S. Fox, Molten Salt Corrosion of SiC and Si3N4, 1988. https://ntrs.nasa.gov/archive/nasa/casi.ntrs.nasa.gov/19890002541.pdf.

[8] J. Kim, M.G. Dunn, A.J. Baran, D.P. Wade, E.L. Tremba, Deposition of Volcanic Materials in the Hot Sections of Two Gas Turbine Engines, in: Vol. 3 Coal, Biomass Altern. Fuels; Combust. Fuels; Oil Gas Appl. Cycle Innov., American Society of Mechanical Engineers, 1992: pp. 641–651. https://doi.org/10.1115/92-GT-219.

[9] N. Al Nasiri, N. Patra, N. Ni, D.D. Jayaseelan, W.E. Lee, Oxidation behaviour of SiC/SiC ceramic matrix composites in air, J. Eur. Ceram. Soc. 36 (2016) 3293–3302. https://doi.org/10.1016/j.jeurceramsoc.2016.05.051.

[10] E.J. Opila, R.E. Hann, Paralinear Oxidation of CVD SiC in Water Vapor, J. Am. Ceram. Soc. 80 (1997) 197–205. https://doi.org/10.1111/j.1151-2916.1997.tb02810.x.

[11] N.S. Jacobson, Corrosion of Silicon-Based Ceramics in Combustion Environments, J. Am. Ceram. Soc. 76 (1993) 3–28. https://doi.org/10.1111/j.1151-2916.1993.tb03684.x.

[12] E.J. Opila, Variation of the Oxidation Rate of Silicon Carbide with Water-Vapor Pressure, J. Am.

Ceram. Soc. 82 (1999) 625–636. https://doi.org/10.1111/j.1151-2916.1999.tb01810.x.

[13]  E.J. Opila, Oxidation Kinetics of Chemically Vapor-Deposited Silicon Carbide in Wet Oxygen, J. Am. Ceram. Soc. 77 (1994) 730–736. https://doi.org/10.1111/j.1151-2916.1994.tb05357.x.

[14]  K. Kane, E. Garcia, P. Stack, M. Lance, C. Parker, S. Sampath, B.A. Pint, Evaluating steam oxidation kinetics of environmental barrier coatings, J. Am. Ceram. Soc. 105 (2022) 590–605. https://doi.org/10.1111/jace.18093.

[15]  E.J. Opila, D.S. Fox, N.S. Jacobson, Mass Spectrometric Identification of Si-O-H(g) Species from the Reaction of Silica with Water Vapor at Atmospheric Pressure, J. Am. Ceram. Soc. 80 (2005) 1009–1012. https://doi.org/10.1111/j.1151-2916.1997.tb02935.x.

[16]  S.L. dos Santos e Lucato, O.H. Sudre, D.B. Marshall, A Method for Assessing Reactions of Water Vapor with Materials in High-Speed, High-Temperature Flow, J. Am. Ceram. Soc. 94 (2011) s186–s195. https://doi.org/10.1111/j.1551-2916.2011.04556.x.

[17]  E.J. Opila, J.L. Smialek, R.C. Robinson, D.S. Fox, N.S. Jacobson, SiC Recession Caused by SiO2 Scale Volatility under Combustion Conditions: II, Thermodynamics and Gaseous-Diffusion Model, J. Am. Ceram. Soc. 82 (1999) 1826–1834. https://doi.org/10.1111/j.1151-2916.1999.tb02005.x.

[18]  D. Tejero-Martin, C. Bennett, T. Hussain, A review on environmental barrier coatings: History, current state of the art and future developments, J. Eur. Ceram. Soc. 41 (2021) 1747–1768. https://doi.org/10.1016/j.jeurceramsoc.2020.10.057.

[19]  A.J. Fernández-Carrión, M. Allix, A.I. Becerro, Thermal Expansion of Rare-Earth Pyrosilicates, J. Am. Ceram. Soc. 96 (2013) 2298–2305. https://doi.org/10.1111/jace.12388.

[20]  Y. Xu, X. Hu, F. Xu, K. Li, Rare earth silicate environmental barrier coatings: Present status and prospective, Ceram. Int. 43 (2017) 5847–5855. https://doi.org/10.1016/j.ceramint.2017.01.153.

[21]  J. Felsche, The crystal chemistry of the rare-earth silicates, in: 1973: pp. 99–197. https://doi.org/10.1007/3-540-06125-8_3.

[22]  L.R. Turcer, N.P. Padture, Towards multifunctional thermal environmental barrier coatings (TEBCs) based on rare-earth pyrosilicate solid-solution ceramics, Scr. Mater. 154 (2018) 111–

117. https://doi.org/10.1016/j.scriptamat.2018.05.032.

[23] K.A. Kane, E. Garcia, S. Uwanyuze, M. Lance, K.A. Unocic, S. Sampath, B.A. Pint, Steam oxidation of ytterbium disilicate environmental barrier coatings with and without a silicon bond coat, J. Am. Ceram. Soc. 104 (2021) 2285–2300. https://doi.org/10.1111/jace.17650.

[24] N. Maier, K.G. Nickel, G. Rixecker, High temperature water vapour corrosion of rare earth disilicates (Y,Yb,Lu)2Si2O7 in the presence of Al(OH)3 impurities, J. Eur. Ceram. Soc. 27 (2007) 2705–2713. https://doi.org/10.1016/j.jeurceramsoc.2006.09.013.

[25] N. Rohbeck, P. Morrell, P. Xiao, Degradation of ytterbium disilicate environmental barrier coatings in high temperature steam atmosphere, J. Eur. Ceram. Soc. 39 (2019) 3153–3163. https://doi.org/10.1016/j.jeurceramsoc.2019.04.034.

[26] M. Ridley, E.J. Opila, Thermochemical stability and microstructural evolution of Yb2Si2O7 in high-velocity high-temperature water vapor, J. Eur. Ceram. Soc. 41 (2021) 3141–3149. https://doi.org/10.1016/j.jeurceramsoc.2020.05.071.

[27] S. Ueno, D.D. Jayaseelan, T. Ohji, Development of Oxide-Based EBC for Silicon Nitride, Int. J. Appl. Ceram. Technol. 1 (2004) 362–373. https://doi.org/10.1111/j.1744-7402.2004.tb00187.x.

[28] K.N. Lee, D.S. Fox, N.P. Bansal, Rare earth silicate environmental barrier coatings for SiC/SiC composites and Si3N4 ceramics, J. Eur. Ceram. Soc. 25 (2005) 1705–1715. https://doi.org/10.1016/j.jeurceramsoc.2004.12.013.

[29] G.C.C. Costa, N.S. Jacobson, Mass spectrometric measurements of the silica activity in the Yb2O3–SiO2 system and implications to assess the degradation of silicate-based coatings in combustion environments, J. Eur. Ceram. Soc. 35 (2015) 4259–4267. https://doi.org/10.1016/j.jeurceramsoc.2015.07.019.

[30] N. Al Nasiri, N. Patra, D.D. Jayaseelan, W.E. Lee, Water vapour corrosion of rare earth monosilicates for environmental barrier coating application, Ceram. Int. 43 (2017) 7393–7400. https://doi.org/10.1016/j.ceramint.2017.02.123.

[31] S. Ueno, D.D. Jayaseelan, T. Ohji, Comparison of water vapor corrosion behavior of silicon nitride with various EBC layers, J. Ceram. Process. Res. 5 (2004) 355–359.

[32]   E. Bakan, Y.J. Sohn, W. Kunz, H. Klemm, R. Vaßen, Effect of processing on high-velocity water vapor recession behavior of Yb-silicate environmental barrier coatings, J. Eur. Ceram. Soc. 39 (2019) 1507–1513. https://doi.org/10.1016/j.jeurceramsoc.2018.11.048.

[33]   E. Bakan, M. Kindelmann, W. Kunz, H. Klemm, R. Vaßen, High-velocity water vapor corrosion of Yb-silicate: Sprayed vs. sintered body, Scr. Mater. 178 (2020) 468–471. https://doi.org/10.1016/j.scriptamat.2019.12.019.

[34]   D. Qin, Y. Niu, H. Li, X. Zhong, X. Zheng, J. Sun, Fabrication and characterization of Yb2Si2O7-based composites as novel abradable sealing coatings, Ceram. Int. 47 (2021) 23153–23161. https://doi.org/10.1016/j.ceramint.2021.05.029.

[35]   P. Scardi, M. Leoni, Whole powder pattern modelling, Acta Crystallogr. Sect. A Found. Crystallogr. 58 (2002) 190–200. https://doi.org/10.1107/S0108767301021298.

[36]   J. Schindelin, I. Arganda-Carreras, E. Frise, V. Kaynig, M. Longair, T. Pietzsch, S. Preibisch, C. Rueden, S. Saalfeld, B. Schmid, J.-Y. Tinevez, D.J. White, V. Hartenstein, K. Eliceiri, P. Tomancak, A. Cardona, Fiji: an open-source platform for biological-image analysis, Nat. Methods. 9 (2012) 676–682. https://doi.org/10.1038/nmeth.2019.

[37]   E. Garcia, O. Sotelo-Mazon, C.A. Poblano-Salas, G. Trapaga, S. Sampath, Characterization of Yb2Si2O7–Yb2SiO5 composite environmental barrier coatings resultant from in situ plasma spray processing, Ceram. Int. 46 (2020) 21328–21335. https://doi.org/10.1016/j.ceramint.2020.05.228.

[38]   S. Sampath, H. Herman, Rapid solidification and microstructure development during plasma spray deposition, J. Therm. Spray Technol. 5 (1996) 445–456. https://doi.org/10.1007/BF02645275.

[39]   B.T. Richards, K.A. Young, F. De Francqueville, S. Sehr, M.R. Begley, H.N.G. Wadley, Response of ytterbium disilicate-silicon environmental barrier coatings to thermal cycling in water vapor, Acta Mater. 106 (2016) 1–14. https://doi.org/10.1016/j.actamat.2015.12.053.

[40]   S. Ueno, T. Ohji, H.-T. Lin, Recession behavior of Yb2Si2O7 phase under high speed steam jet at high temperatures, Corros. Sci. 50 (2008) 178–182. https://doi.org/10.1016/j.corsci.2007.06.014.

[41] E. Bakan, D.E. Mack, S. Lobe, D. Koch, R. Vaßen, An investigation on burner rig testing of environmental barrier coatings for aerospace applications, J. Eur. Ceram. Soc. 40 (2020) 6236–6240. https://doi.org/10.1016/j.jeurceramsoc.2020.06.016.

[42] Y. Murakami, H. Yamamoto, Phase Equilibria and Properties of Glasses in the Al2O3-Yb2O3-SiO2 System, J. Ceram. Soc. Japan. 101 (1993) 1101–1106. https://doi.org/10.2109/jcersj.101.1101.

[43] L.R. Turcer, A.R. Krause, H.F. Garces, L. Zhang, N.P. Padture, Environmental-barrier coating ceramics for resistance against attack by molten calcia-magnesia-aluminosilicate (CMAS) glass: Part II, β-Yb2Si2O7 and β-Sc2Si2O7, J. Eur. Ceram. Soc. 38 (2018) 3914–3924. https://doi.org/10.1016/j.jeurceramsoc.2018.03.010.

[44] J.L. Stokes, B.J. Harder, V.L. Wiesner, D.E. Wolfe, High-Temperature thermochemical interactions of molten silicates with Yb2Si2O7 and Y2Si2O7 environmental barrier coating materials, J. Eur. Ceram. Soc. 39 (2019) 5059–5067. https://doi.org/10.1016/j.jeurceramsoc.2019.06.051.

[45] J. Liu, L. Zhang, Q. Liu, L. Cheng, Y. Wang, Calcium–magnesium–aluminosilicate corrosion behaviors of rare-earth disilicates at 1400°C, J. Eur. Ceram. Soc. 33 (2013) 3419–3428. https://doi.org/10.1016/j.jeurceramsoc.2013.05.030.

[46] F. Stolzenburg, M.T. Johnson, K.N. Lee, N.S. Jacobson, K.T. Faber, The interaction of calcium–magnesium–aluminosilicate with ytterbium silicate environmental barrier materials, Surf. Coatings Technol. 284 (2015) 44–50. https://doi.org/10.1016/j.surfcoat.2015.08.069.